# A historical perspective on developing foundations for iInfo™ information systems:
*iConsult*™ *and iEntertain*™ apps using *iInfo*™ information integration for *iOrgs*™ information systems

Carl Hewitt

http://carlhewitt.info

*This paper is dedicated to J.C.R. "Lick" Licklider and Marvin Minsky.*

## Abstract

Technology now at hand can integrate all kinds of digital information for individuals, groups, and organizations so their information usefully links together. iInfo™ information integration works by making connections including examples like the following:

- A statistical connection between "being in a traffic jam" and "driving in downtown Trenton between 5PM and 6PM on a weekday."
- A terminological connection between "MSR" and "Microsoft Research."
- A causal connection between "joining a group" and "being a member of the group."
- A syntactic connection between "a pin dropped" and "a dropped pin."
- A biological connection between "a dolphin" and "a mammal".
- A demographic connection between "undocumented residents of California" and "7% of the population of California."
- A geographical connection between "Leeds" and "England."
- A temporal connection between "turning on a computer" and "joining an on-line discussion."

By making these connections, iInfo offers tremendous value for individuals, families, groups, and organizations in making more effective use of information technology.

In practice, integrated information is invariably pervasively inconsistent. Therefore iInfo must be able to make connections even in the face of inconsistency. The business of iInfo is not to make difficult decisions like deciding the ultimate truth or probability of propositions. Instead it provides means for processing information and carefully recording its provenance including arguments (including arguments about arguments) for and against propositions that is used by iConsult<sup>TM</sup> and iEntertain<sup>TM</sup> apps in iOrgs<sup>TM</sup> information systems.

A historical perspective on the above questions is highly pertinent to the current quest to develop foundations for privacy-friendly client-cloud computing.

# Contents

# Privacy-friendly client-cloud computing

In client-cloud computing, information is permanently stored in datacenters on the Internet and cached temporarily on clients that range from single chip sensors, handhelds, notebooks, desktops, and entertainment centers to huge data centers. (Even data centers often cache their information to guard against geographical disaster.) Client-cloud computing will provide new capabilities including the following:

- maintaining the privacy of client information by storing it on datacenters encrypted so that it can be decrypted only by using the client's private key.
- allowing greater integration of information obtained from datacenters of competing integrators (Google, Microsoft, Facebook, etc.). A consequence is that clients can become a monetization platform providing better advertising relevance and targeting without exposing client privacy. Also datacenters of different integrators can treat each other symmetrically.
- allowing clients to provide convenient ways to share their information. For example, medical information will be sent out that can be decrypted by medical providers. Also, appropriate information feeds will be provided for social sharing with family, friends, colleagues, and followers.

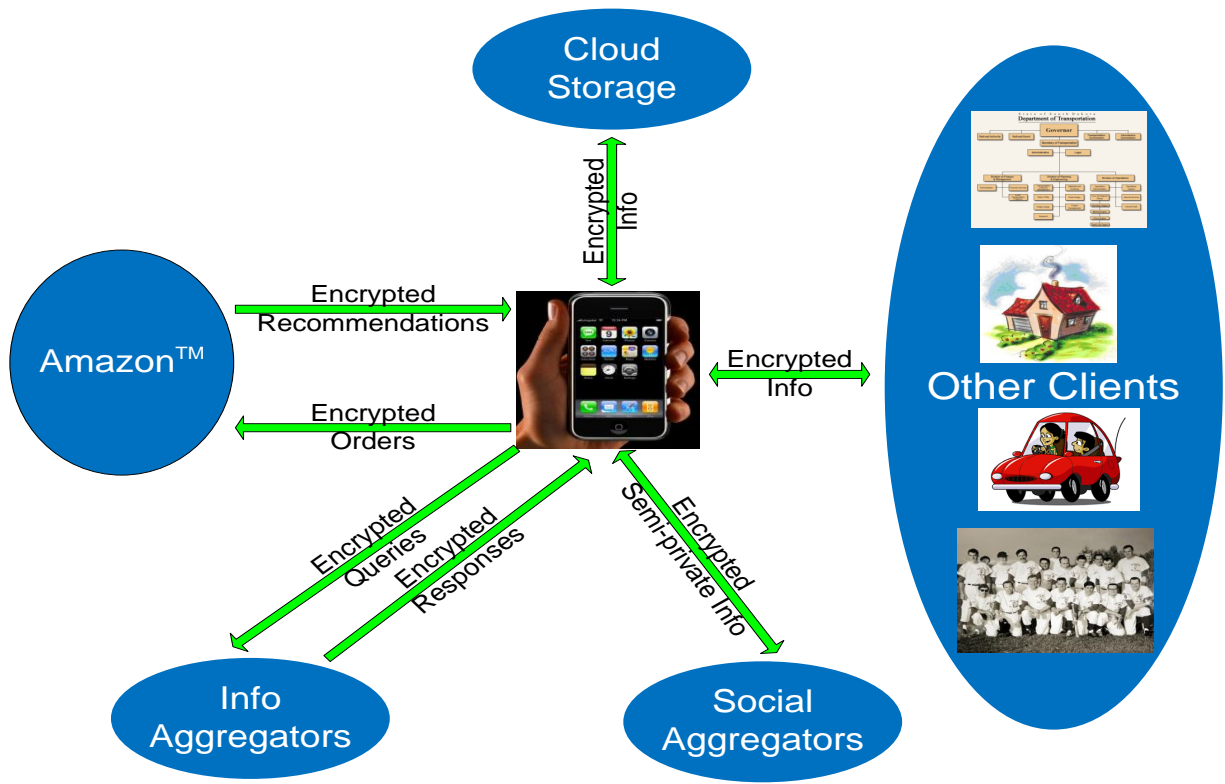

Privacy-friendly Client-cloud Computing

*Information Integration using Client Equipment*

Many consumers value their privacy and do not want to have their intimate personal information stored unencrypted in aggregator datacenters where it can be subpoenaed and observed by an aggregator's employees. Datacenter integration will face the challenge of emerging privacy-friendly competitors that will perform information integration using client equipment instead of cloud datacenters. As explained below, these competitors will have important advantages over datacenter integration including increased revenue from better targeted advertising, lower operational costs, fewer requirements for government regulation, and greater consumer and advertiser satisfaction.

*Lower Costs*

Information integration can be computationally intensive. It's less expensive for competitors to perform integration using client equipment than datacenter integration. Also, competitors using client information integration will have lower communications costs than datacenter integration because communication by clients with datacenters will be less necessary.

*Faster Response*

Because it is less necessary to communicate with datacenters, competitors using client information integration can provide faster response than datacenter integration because most needed information will already be cached in the client equipment.

*Less Regulation*

By performing information integration in clients, competitors can store consumer information in datacenters encrypted so that it can be decrypted only using the client's private key. In this way, there will be fewer requirements for regulating these competitors than those using datacenter integration because they will have less intimate personal information in their datacenters.

Of course, competitors that perform information integration on client equipment will provide convenient ways for consumers to share their information. For example, personal medical information will be sent out that can be decrypted by their medical providers. Also, appropriate information feeds will be provided for social sharing with family, friends, colleagues, and followers.

In summary, competitors can make more money with greater consumer and advertiser satisfaction by integrating information using client equipment than datacenter integration. Also, there will fewer requirements for the government to regulate them.

*iInfo*™ *Information Integration*

Technology now at hand can integrate all kinds of digital information for individuals, groups, and organizations so their information usefully links together. This integration can include calendars and to-do lists, communications (including email, SMS, Twitter, Facebook), presence information (including who else is in the neighborhood), physical (including GPS recordings), psychological (including facial expression, heart rate, voice stress) and social (including family, friends, team mates, and colleagues), maps (including firms, points of interest, traffic, parking, and weather), events (including alerts and status), documents (including presentations, spreadsheets, proposals, job applications, health records, photos, videos, gift lists, memos, purchasing, contracts, articles), contacts (including social graphs and reputation), purchasing information (including store purchases, web purchases, GPS and phone records, and buying and travel habits), government information (including licenses, taxes, and rulings), and search results (including rankings and ratings).

iInfo works by making connections including examples like the following:
- A statistical connection between "being in a traffic jam" and "driving in downtown Trenton between 5PM and 6PM on a weekday."
- A terminological connection between "MSR" and "Microsoft Research."
- A causal connection between "joining a group" and "being a member of the group."
- A syntactic connection between "a pin dropped" and "a dropped pin."
- A biological connection between "a dolphin" and "a mammal".
- A demographic connection between "undocumented residents of California" and "7% of the population of California."
- A geographical connection between "Leeds" and "England."
- A temporal connection between "turning on a computer" and "joining an on-line discussion."

By making these connections iInfo offers tremendous value for individuals, families, groups, and organizations in making more effective use of information technology.

In practice integrated information is invariably inconsistent.[i] Therefore iInfo must be able to make connections even in the face of inconsistency.[ii] The business of iInfo is not to make difficult decisions like deciding the ultimate truth or probability of propositions. Instead it provides means for processing information and carefully recording its provenance including arguments (including arguments about arguments) for and against propositions.

iInfo can make use of the following information system principles:
- *Monotonicity*. *Information is collected and indexed.*
- *Concurrency*: *Work proceeds interactively and concurrently, overlapping in time.*
- *Commutativity*: *Information can be used regardless of whether it initiates new work or become relevant to ongoing work.*
- *Sponsorship*: *Sponsors provide resources for computation, i.e., processing, storage, and communications.*
- *Pluralism*: *Information is heterogeneous, overlapping and often inconsistent. There is no central arbiter of truth*
- *Provenance*: *The provenance of information is carefully tracked and recorded*

There is tremendous value in integrating the above kinds of information. As an example, consider the operation of iInfo with an on-line forum [Google 2009; Epipheo Studios 2009] like the following:

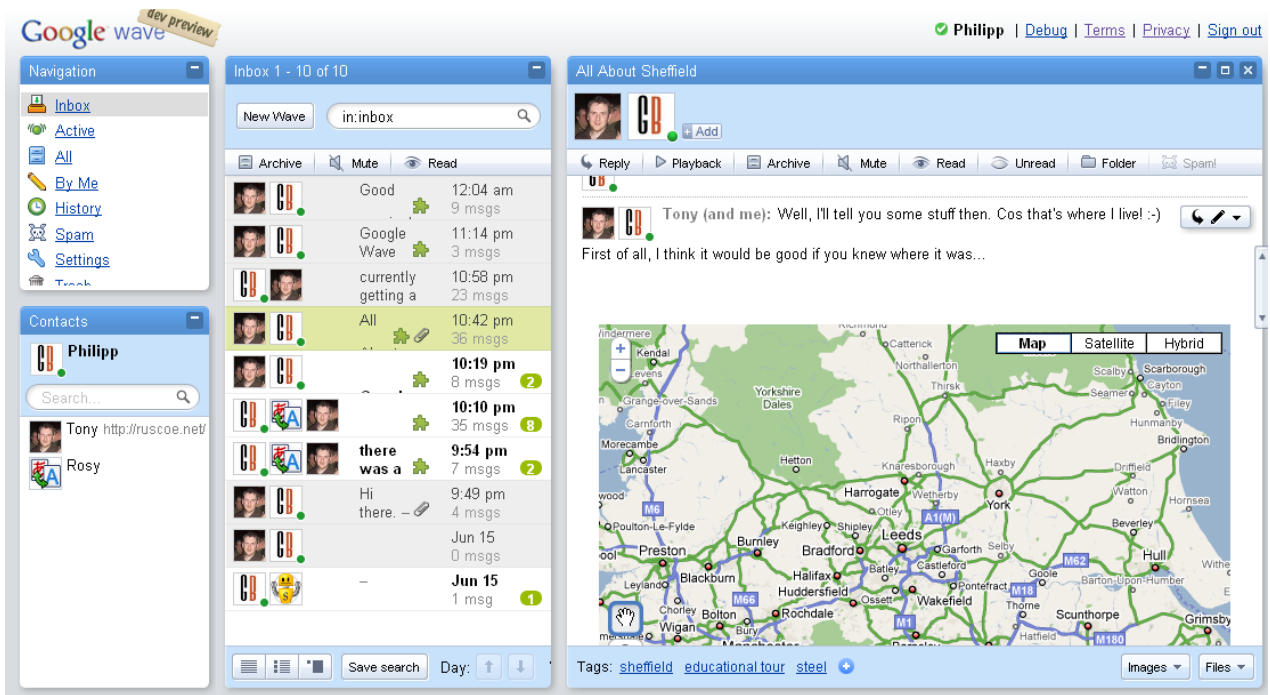
An on-line forum

## iPhrases™ Natural Language

iPhrases[iii] can be used to interpret natural language such as the following:
- *"Make a table of when each user from Cambridge Microsoft Research joined this discussion."*
- *"Drop a pin on the map where Tony's mom lives."*
- *"When is Rosy returning from Leeds after she sees Tony?"*.

## iOrgs™ Information Systems

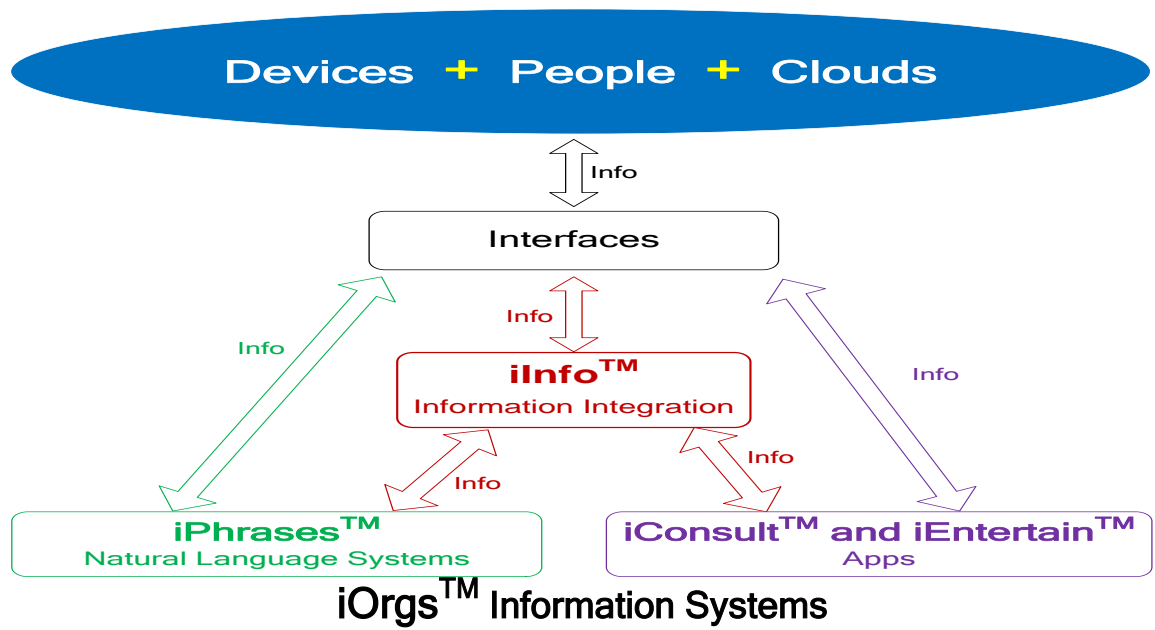
iOrgs™ Information Systems

iOrgs provide the capability for clients to manage technology for integrating their information. Users act as the ultimate managers of their iOrgs, which are accountable to the users for their actions. Implementing iOrgs on clients' equipment can provide the right kind of balance between consumers, merchants, and aggregators like Google, Microsoft, Facebook, etc.[iv]

*Mediated Communication[v]*

Mediated communication is "the process by which communication is mediated through intermediaries."

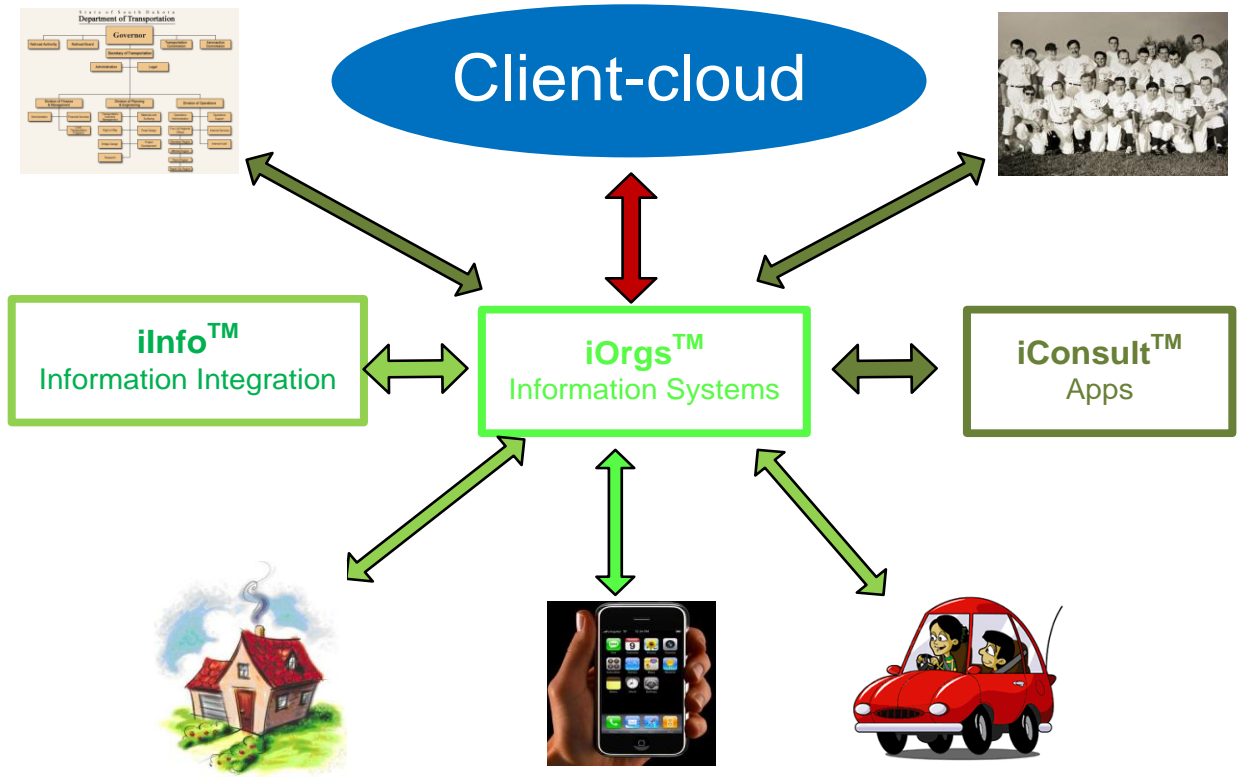

**Mediated Communication**

In the future, persuaders, for the most part, will not be able to interact directly with people. Instead the interactions will be mediated through their iOrgs. Instead of being able to directly communicate with people, persuaders' interactions will be mediated by peoples' Information Technology. Commercial and political advertising will be processed and organized by iOrgs that will solicit competitive views and arguments. Consequently, technologies for persuasion will be completely revolutionized.

Entertainment is a large loophole in mediated communication. In order to be entertained, people are willing to give their undivided attention.

## *iConsult*[TM] *and iEntertain*[TM] *Apps*

Third party applications that can provide entertainment and act as consultants to iOrgs.

Consider the following example in his address to the 2009 Democratic Caucus, Steve Ballmer said:
> *Instead of telling my secretary to get me ready for my trip to the House Democratic Caucus, I'll just type it in or speak it to my computer.* **It can look up, it turns out, who you all** [the caucus] ***are, and where you're all from, and it's got all–it's all out there. We just need to automate it in ways that real people can get access to information.*** [Ballmer 2009] (emphasis added)

As another example, consider the following task for iInfo:
- ***Working for ABC Corp, your task is to organize a joint sales conference with your partner XYZ Corp.***
- *It will include approximately 60 regional sales managers from both companies including international (visas will be required). The conference will be for 2 days in the summer of 2009 in the Western US at a scenic location near golf links.*
- *There will be an awards banquet (with individually engraved plaques for awardees). The sales VPs of both companies must attend. The air and car rental travel of participants should be coordinated to maximize interaction.*
- *The conference budget for ABC Corp is $60K.*
- ***You need to prepare a detailed proposal for the sales VPs of both companies in 1 week's time!***

Today's computer systems offer little more than "copy and paste" to aid integration for the above tasks. iInfo enable iConsult and iEntertain apps with much greater functionality.

Additional iInfo apps can include:
- Email sorting and summarization
- Product reviewing and recommending
- Travel planning (including coordinating multiple travelers, destinations and preferences)
- Personal Scheduling
- Social recommendations for products and services

## *Interfaces*

The iOrgs platform provides interfaces to the outside including media (sound and display), sensors (cameras, movement, touch, physiological), home (security, energy, kitchen), and Internet (peer-to-peer, social sharing, feeds).

*Paradigm Shift from Personal Computing to Information Integration*
It's instructive to compare the Information Integration paradigm shift with the previous Personal Computer paradigm shift in terms of user interfaces, processing, network security, and operations:

|  | **Personal Computing** | **Information Integration** |
|---|---|---|
| **Search and Discovery:** | keywords | iPhrases[TM] Natural language |
| **Network Security:** | firewall | client-cloud encryption |
| **Processing:** | microprocessor | many-core |
| **Information Integration Robustness** | inconsistency denial | inconsistency robustness |
| **Robustness** | app termination | rapid recovery |
| **Applications** | attention controlling | Mediated iConsult[TM] and iEntertain[TM] Apps |
| **Information Integration:** | file system | iInfo[TM] Information Integration |
| **Operational Control:** | operating system | iOrgs[TM] Information Systems |
| **User Attention:** | application centered | mediated |

Paradigm Shift

**What went wrong:**
1. Numerous kinds of information of great importance to clients were stored separately with little more than "*copy, cut, and paste*" for integration.
2. Aggregators (Google, Microsoft, Yahoo, Facebook, *etc*.) stored client information in their datacenters in such a way that there were increasingly subject to heavy government regulation.
3. Applications control user attention

**What is being done about it:**
1. iInfo is being developed to integrate information from diverse sources thereby enormously increasing the value of information technology.
2. iOrgs can store client information encrypted in aggregator datacenters that can be unencrypted only by using a client's private key.
3. User attention mediated by iOrgs.

# We are in the midst of a paradigm shift from *"inconsistency denial"* to *practical information integration*

## Development of iOrgs
The continuing development of Moore's Law has imposed increasing problems of scalability that have greatly impacted the development of software engineering.

### *Mental Agents*[vi]
A Mental Agent is defined behaviorally as cognitively operating like a human. The paradigm is deeply and pervasively psychological.[Wang and Laird 2006; Minsky 2006] The most popular kind of mental agent can be characterized as BDIA: beliefs, desires (goals), intentions (plans), and affect (emotions). It has moved beyond its original sequential conceptualization by introducing parallelism, which can be used for low-level input–output (vision, for example), (associative) memory operations, and other basic operations as performed by the parts of the brain. Yet, none of these changes the mental agent paradigm, which draws its fundamental strength from staying close to the mental operations of a single person.

The development of mental agents has continued steadily since the earliest days of artificial intelligence, and researchers have realized impressive achievements (see the Summer 2006 issue of AI Magazine for some examples). That said, progress in using mental agents as a foundation for software applications has been frustratingly slow. Impressive demonstrations of mental agents' capabilities in some application areas have repeatedly failed to garner widespread commercial adoption. Nevertheless, researchers continue the quest to develop mental agent frameworks for software systems. Expressions of confidence and hopes for the future have long been regular features of conferences.[Benfield, Hendrickson, and Galanti 2006] Researchers have been heartened because no convincing principled arguments have shown it impossible; indeed, human behavior presents a kind of existence proof that something like the mental agent paradigm can be made to work. Moreover, the community has evolved and gained insights into multi-agent systems.

The perfect disruption is causing mental agents to lose ground. First, many-core architectures pose a challenge because the information processing of a computer is no longer at all like the information processing of a person:
- Using human-like mental operations becomes an increasing bottleneck as the number of cores increase because the cores perform independent tasks.
- Using a human-like input/output system becomes an increasing bottleneck as the number of interconnections increase because the wires carry independent messages.

According to a published consensus of researchers, a software agent is basically a mental agent adapted for software engineering.[Huhns, *et. al.* 2005] More general conceptions have been attempted without success. As Charles Petrie noted, for example, "*some have tried to offer the general definition of agents as someone or something that acts on one's behalf, but that seems to cover all of computers and software*."[Petrie 1996].

Many artificial intelligence researchers have long presupposed that agents are a principal subject of their field. This is especially poignant for the autonomous agents and multi-agent systems (AAMAS) community, which includes the term "agent" in its name twice. AAMAS is a vibrant community whose members are performing exciting and important research, but its conceptual foundations are badly in need of reformulation because of lack of success using Mental Agents in large software systems.

## *In the organization lies the power*

iOrgs[vii] has the goal of becoming an effective readily understood approach for addressing scalability issues in Software Engineering. The paradigm takes its inspiration from human organizations. iOrgs provide a framework for addressing issues of hierarchy, authority, accountability, scalability, and robustness using methods that are analogous to human organizations. Because humans are very familiar with the principles, methods, and practices of human organizations, they can transfer this knowledge and experience to iOrgs. iOrgs achieve scalability by mirroring human organizational structure. For example an iOrg can have sub-organizations specialized by areas such as sales, production, and so forth. Authority is delegated down the organizational structure and when necessary issues are escalated upward. Authority requires accountability for its use including record keeping and periodic reports. Management is in large part the art of reconciling authority and accountability.

iOrgs are structured around *organizational commitment* defined as information pledged constituting an alliance to go forward. For example, iOrgs can use contracts to formalize their mutual commitments to fulfill specified obligations to each other. Yet, manifestations of information pledged will often be inconsistent. Any given agreement might be internally inconsistent, or two agreements in force at one time could contradict each other. Issues that arise from such inconsistencies can be negotiated among iOrgs.

## *Paradigm Shift from Mental Agents to iOrgs*

Despite many years of trying, none of the software agent development systems for large-scale Internet applications have had any significant commercial success. Organizational computing is trumping mental agents for implementing large-scale Internet systems. No software agent architectures can compete with iOrgs in understandability, manageability, and scalability.[viii] As [Petrie 2000] predicted the old agent technology has essentially disappeared from large-scale software systems. Consequently, a conundrum is emerging, such that researchers must choose whether to
- stay the current mental agents course despite the paradigm switch from mental agents to iOrgs, or
- change course and adopt the iOrgs paradigm as fundamental, thus begging the question, "Where are the agents?"

The mental agent paradigm might increasingly be used in avatars (both human-like and animal-like) and cognitive models [Norling and Ritter 2004] of individual humans, but operational implementations will require iOrgs, just as all large software systems will. The original conception of the development of mental agents is thus turned upside down: instead of iOrgs being implemented using mental agents, humans will be simulated and avatars will be implemented using iOrgs! The same applies to communities of agents in which multitudes of agents communicate [Chainbi 2003].

## *Rapid Recovery*

*Rapid Recovery* is a computing paradigm being developed in contrast with the traditional *Inconsistency Denial* paradigm.

Digital data is fragile. It often doesn't take much to make it unrecoverable. Consequently, we adopt the following principle:

> **All data is cached data; however, sometimes there is only one copy**.

For example, consider a cloud blob storage service that stores and retrieves digital artifacts (called blobs). Amazon Dynamo [DeCandia, *et al.* 2007] and Tahoe [Wilcox-O'Hearn and Warner 2008] developed highly available blob storage services that could be improved in the following ways:

- Making storage receipt-based instead of key-based. In receipt-based storage a receipt is provided for each instance of the deposit of a blob, a familiar business model to customers. Receipt-based storage can be more efficiently implemented than key-based because it does not require global co-ordination of keys.
- Making each deposit of a blob under a Service Level Agreement (SLA) that can be of various kinds including the following:
    - rent per time period
    - incremental charge for retrieving the blob
    - drop-off changes for retrieving the blob at a place that is geographically distant from where it was stored
    - incremental charge for replacing the blob with a new version and issuing a replacement receipt. The replacement can optionally be specified as an incremental difference of the blob being replaced in order to save on storage and communications.

- variable charging for availability and reliability
- requiring that blob retrieval in addition to requiring a receipt must also be used by an authenticated iOrg.[ix]
* Providing a clean abstraction for high availability in retrieving blobs. A request to retrieve the blob for a receipt should either return the blob or throw an exception if the SLA specified when depositing the blob cannot be met. However, the exception can provide partial information and the ability to later receive additional information. For example, the exception can include a list, each element of which is an alternative previous version of the blob together with the receipt that was provided when it was stored.

In contrast, *Rapid Recovery* can be compared with *Eventual Consistency* [Vogels 2007]:

*The storage system guarantees that if no new updates are made to the object eventually (after the inconsistency window closes) all accesses will return the last updated value.*

*Rapid Recovery* differs from *Eventual Consistency* as follows:
1. In response to a request to retrieve a blob for a receipt, the blob storage system may respond that, unfortunately, all versions of the blob have been irretrievably lost. In which case, (monetary) compensation may be owed in accordance with the SLA of the receipt.
2. It may not be possible to retrieve the latest version of a blob using the receipt that was proved when the version was stored. Only older versions of the blob might be available.
3. Recovery information can be provided in the exception thrown by a request that does not meet its SLA. For example, the exception can include an estimate as to when a better response to the request might be available.
4. A request can be made that better responses be sent as they become available; *i.e.,* to provide rapid recovery.

**What went wrong:**
1. The Mental Agent paradigm turned out to be too restrictive because of the "perfect disruption" involving:
    a. *Hardware*. Many-core architecture
    b. *Software*. iInfo information integration
    c. *Applications*. iConsult apps for iOrgs information systems
2. Traditional data parallel systems (e.g. MapReduce [Dean and Ghemwat 1994] and Dryad [Michael Isard, et. al. 2007]) lack generality.[Turing 1949, McCarthy 1963, Floyd 1967, and Hoare 2003] proposed using classical mathematical logic to prove that programs were consistent with their specifications. However, as systems grew larger this became infeasible [Cusumano and Selby 1995, Rosenberg 2007].
3. Dynamo and Tahoe developed highly available cloud storage services that although practically useful for what they were designed did not implement Rapid Recovery (*i.e.* functionality to return improved responses to requests as they became available).

**What was done about it:**
1. iOrgs were developed to meet the requirements of the perfect disruption.
2. The iOrgs paradigm is strictly more general than data parallelism (*e.g.* MapReduce and Dryad).
3. An improved abstraction is being developed for Rapid Recovery cloud storage services.[x]

# Direct Logic™: In the argumentation lies the knowledge.

Direct Logic [Hewitt 2008f] is the minimal fix to classical mathematical logic and statistical probability (fuzzy) inference that meets the requirements of large-scale Internet applications (including sense making for natural language) by addressing the following issues: inconsistency robustness, contrapositive inference bug, and direct argumentation.[xi]

## *Pervasive Inconsistency*

The development of large software systems and the extreme dependence of our society on these systems have introduced new phenomena. These systems have pervasive inconsistencies among and within the following:
- *Use cases* that express how systems can be used and tested in practice
- *Documentation* that expresses over-arching justification for systems and their technologies
- *Code* that expresses implementations of systems

Different communities are responsible for constructing, evolving, justifying and maintaining documentation, use cases, and code for large, human-interaction, software systems. In specific cases any one consideration can trump the others. Sometimes debates over inconsistencies among the parts can become quite heated, *e.g.,* between vendors. ***In the long run, after difficult negotiations, in large software systems, use cases, documentation, and code all change to produce systems with new inconsistencies. However, no one knows what they are or where they are located!***

Furthermore there is no evident way to divide up the code, documentation, and use cases into meaningful, consistent microtheories for human-computer interaction. ***Organizations such as Microsoft, the US government, and IBM have tens of thousands of employees pouring over hundreds of millions of lines of documentation, code, and use cases attempting to cope. In the course of time almost all of this code will interoperate using Web Services. A large software system is never done*** [Rosenberg 2007]**.**

The thinking in almost all scientific and engineering work has been that models (also called theories or microtheories) should be internally consistent, although they could be inconsistent with each other.

Consistency testing is recursively undecidable even in first order logic. Because of this difficulty, it is usually not known whether or not large theories of practical domains are consistent. In practice, the information in large software projects and information on the Internet is invariably inconsistent.

## *Direct Inference*

Direct Logic is based on direct inference[xii] to more directly infer conclusions from premises.

Direct inference is used in to directly infer conclusions from premises. For example, suppose that we have

**A1.** Observe[WeekdayAt5PM] $\vdash_{Boston}$ TrafficJam[xiii]

**A2.** $\vdash_{Boston}$ ¬TrafficJam[xiv]

In classical logic, ¬ Observe [WeekdayAt5PM] is inferred in theory $\mathcal{Boston}$ from **A1** and **A2** above.[xv] But in Direct Logic contraposition does not hold for inference. Consequently, ***direct inference comes into play even in the absence of overt inconsistency.***[xvi]

In this respect, the Deduction Theorem of logic plays a crucial role in relating logical implication to computation. The *Classical Deduction Theorem* can be stated as follows: $(\vdash (\Psi \Rightarrow \Phi)) \Leftrightarrow (\Psi \vdash \Phi)$ stating that $\Psi \Rightarrow \Phi$ can be inferred if and only if $\Phi$ can be inferred from $\Psi$. Thus procedures can search for an inference of the implication $\Psi \Rightarrow \Phi$ by simply searching for an inference of $\Phi$ from $\Psi$. *However, the Classical Deduction Theorem is not valid for Direct Logic.*

*Two-way Deduction Theorem*
Consequently for Direct Logic, the *Two-Way Deduction Theorem* [Hewitt 2008f] was developed taking the following form:

$$\vdash_T (\Psi \Rightarrow \Phi) \quad \dashv \vdash_T \quad (\Psi \vdash_T \Phi) \wedge (\neg \Phi \vdash_T \neg \Psi)$$

stating that $\Psi \Rightarrow \Phi$ can be inferred in a theory $T$ is provably equivalent to both $\Phi$ can be inferred in $T$ from $\Psi$ *and* $\neg \Psi$ can be inferred in $T$ from $\neg \Phi$. In this way, the Two-Way Deduction Theorem provides an extension of natural deduction for implications in Direct Logic.

Direct inference is reasoning that requires a more direct inferential connection between premises and conclusions than classical logic. For example, in classical logic, (not WeekdayAt5PM) can be inferred from the premises (not TrafficJam) and (WeekdayAt5PM infers TrafficJam). However, direct inference does not sanction concluding (not WeekdayAt5PM) because it might be that there is no traffic jam but it undesirable to infer (not WeekdayAt5PM).[xvii]

In summary, Direct Logic has important advantages over previous proposals (e.g. Relevance Logic) to more directly connect antecedents to consequences in reasoning. These advantages include:
- using natural deduction reasoning
- preserving the standard Boolean equivalences (double negation, De Morgan, etc.)
- being able to more safely reason about the mutually inconsistent data, code, specifications, and test cases of client cloud computing
- having an intuitive deduction theorem that connects logical implication with inference.
- inference in Boolean[xviii] Direct Logic is recursively decidable[xix]

Direct Logic preserves as much of classical logic as possible given that it is based on direct inference.

*Logical Necessity of Inconsistency*
Platonic Ideals were to be perfect, unchanging, and eternal. Beginning with the Hellenistic mathematician Euclid [circa 300BC] in Alexandria, theories were intuitively supposed to be both consistent and complete. However, Gödel [1931] (later generalized by Rosser [1936]) proved that mathematical theories are incomplete, *i.e.*, there are propositions that neither the proposition nor its negation can be inferred. This was accomplished by showing that in each sufficiently strong theory $T$, there is a paradoxical proposition

$\text{Uninferable}_T$ that is logically equivalent to its own uninferability, *i.e.*, $\nvdash_T \text{Uninferable}_T$

To demonstrate the power of Direct Logic, a generalization of the incompleteness theorem was proved without using the assumption of consistency on which Gödel/Rosser had relied for their proofs. Then there was a surprising development: since it turns out that the Gödelian paradoxical proposition $\text{Uninferable}_T$ is self-inferable (*i.e.* $\vdash_T \text{Uninferable}_T$), it follows that every theory in Direct Logic is inconsistent! However, in the context of large software systems, it is not especially bothersome that theories of Direct Logic are inconsistent about $\vdash_T \text{Uninferable}_T$.

According to Hewitt [2008f]:
> This means that the formal concept of TRUTH as developed by Tarski, *et al.* is out the window. At first, TRUTH may seem like a desirable property for propositions in theories for large software systems. However, because a theory $T$ is necessarily inconsistent about $\vdash_T \text{Uninferable}_T$ it is impossible to consistently assign truth values to propositions of $T$. In particular it is impossible to consistently assign a truth value to the proposition $\vdash_T \text{Uninferable}_T$. If the proposition is assigned the value TRUE, then (by the rules for truth values) it must also be assigned FALSE and vice versa. It is not obvious what (if anything) is wrong or how to fix it."

## *What about statistical probability (fuzzy) inference?*

Statistical probabilistic (fuzzy logic) systems are affected follows: Suppose (as above)

$$\vdash_{Boston} P(\text{Observe}[\text{TrafficJam}] \mid \text{WeekdayAt5PM}) = 1^{\text{xx}}$$

$$\vdash_{Boston} P(\text{TrafficJam}) = 0$$

Then

$$\vdash_{Boston} P(\text{Observe}[\text{WeekdayAt5PM}]) = \frac{P(\text{WeekdayAt5PM} \wedge \text{Observe}[\text{TrafficJam}])}{P(\text{TrafficJam} \mid \text{Observe}[\text{WeekdayAt5PM}])} = 0^{\text{xxi}}$$

Thus contraposition is built into probabilistic (fuzzy logic) systems and consequently incorrect information can be generated.

The above example illustrates that the choice of how to incorporate measurements into statistics can effectively determine the model being used. In this particular case, the way that measurements were taken did not happen to take into account things like holidays and severe winter storms This point was largely missed in [Anderson 2008] which stated
> *"Correlation is enough."* **We can stop looking for models. We can analyze the data without hypotheses about what it might show.** *We can throw the numbers into the biggest computing clusters the world has ever seen and let statistical algorithms find patterns where science cannot."* (emphasis added)

October 4, 2010                                                                                                      Page 15 of 37

**What went wrong:**
1. Cyc [Masters and Gűngőrdű 2003] and the specification of OWL[xxii] 2 [Motik, Patel-Schneider and Grau 2008] incorporated the assumption that if a theory[xxiii] is not absolutely consistent then anything and everything can be inferred.[xxiv] Furthermore, OWL 2 lacks support for statistical and probabilistic inference (*e.g.,* see [Neapolitan 2004]).
2. Pure Logic Programming turned out to be too restrictive to handle the information processing for large-scale open concurrent systems.
3. Classical mathematical logic and probabilistic (fuzzy logic) systems made use of indirect inference that made them unsafe for use in reasoning about inconsistent information.
4. The Classical Deduction Theorem (a mainstay principle of Logic Programming) was found not to be valid for theories of Direct Logic.
5. When formalizing reasoning for large software systems, the reasoning process itself produced inconsistencies about certain specialized propositions that make assertions about their own uninferable.
6. Statistical probabilistic (fuzzy logic) systems ran into trouble with incorrect and inconsistent information.

**What is being done about it:**
1. Cyc and the OWL 2 specification need to be updated to incorporate inconsistency robustness [Hewitt 2008f, 2008g] and iPhrases Natural Language [xxv].
2. Less restrictive principles are being developed that generalize/revise principles of Logic Programming based on the Scientific Community Metaphor [Kornfeld and Hewitt 1981]. Moveable Objects [Helland 2007] and iOrgs.
3. Direct Logic was developed for reasoning about the pervasively inconsistent information in the data, code, specifications, and test cases of large software systems.
4. A replacement for the classical Deduction Theorem (the Two-way Deduction Theorem) was developed thereby facilitating Logic Programming using Direct Logic
5. It was decided to live with these inconsistencies because:
   - The inconsistencies about the self-inferability of propositions are irrelevant for large software systems that are chock full of other inconsistencies that do matter.
   - The inconsistencies about self-inferability do no great harm since they have no relevant consequences for large software systems.
6. Direct Logic is tolerant of the inconsistencies in the information that goes into statistical measurements.

## Actors

Several models of nondeterministic computation were developed including the following:

### Petri nets

Prior to the development of the Actor model, Petri nets were widely used to model nondeterminism. However, they were widely acknowledged to have an important limitation: they modeled control flow but not data flow. Consequently they were not readily composable. Another difficulty with Petri nets is simultaneous action. *I.e*., the atomic step of computation in Petri nets is a transition in which tokens simultaneously disappear from the input places of a transition and appear in the output places. The physical basis of using a primitive with this kind of simultaneity seems questionable. Despite these apparent difficulties, Petri nets continue to be a popular approach to modeling nondeterminism, and are still the subject of active research.

### Simula

Simula 1 [Nygaard 1962] pioneered nondeterministic discrete event simulation using a global clock:

*In this early version of Simula a system was modeled by a (fixed) number of "stations", each with a queue of "customers". The stations were the active parts, and each was controlled by a program that could "input" a customer from the station's queue, update variables (global, local in station, and local in customer), and transfer the customer to the queue of another station. Stations could discard customers by not transferring them to another queue, and could generate new customers. They could also wait a given period (in simulated time) before starting the next action. Custom types were declared as data records, without any actions (or procedures) of their own.* [Krogdahl 2003]

Thus at each time step, the program of the next station to be simulated would update the variables.

Kristen Nygaard and Ole-Johan Dahl developed the idea (first described in an IFIP workshop in 1967) of organizing objects into "classes" with "subclasses" that could inherit methods for performing operations from their superclasses. In this way, Simula 67 considerably improved the modularity of nondeterministic discrete event simulations.

According to [Krogdahl 2003]:
> Objects could act as processes that can execute in "quasi-parallel" that is in fact a form of nondeterministic sequential execution in which a simulation is organized as "independent" processes. Classes in Simula 67 have their own procedures that start when an object is generated. However, unlike Algol procedures, objects may choose to temporarily stop their execution and transfer the control to another process. If the control is later given back to the object, it will resume execution where the control last left off. A process will always retain the execution control until it explicitly gives it away. When the execution of an object reaches the end of its statements, it will become "terminated", and can no longer be resumed (but local data and local procedures can still be accessed from outside the object).
>
> The quasi-parallel sequencing is essential for the simulation mechanism. Roughly speaking, it works as follows: When a process has finished the actions to be performed at a certain point in simulated time, it decides when (again in simulated time) it wants the control back, and stores this in a local "next-event-time" variable. It then gives the control to a central "time-manager", which finds the process that is to execute next (the one with the smallest next-event-time), updates the global time variable accordingly, and gives the control to that process.
>
> The idea of this mechanism was to invite the programmer of a simulation program to model the underlying system by a set of processes, each describing some natural sequence of events in that system (e.g. the sequence of events experienced by one car in a traffic simulation).
>
> Note that a process may transfer control to another process even if it is currently inside one or more procedure calls. Thus, each quasi-parallel process will have its own stack of procedure calls, and if it is not executing, its "reactivation point" will reside in the innermost of these calls. Quasi-parallel sequencing is analogous to the notion of co-routines [Conway 1963].

Note that a class operation operated on the global state of the simulation and not just on the local variables of the class in which the operation is defined.[xxvi] Also Simula-67 lacked formal interfaces and instead relied on inheritance from an abstract class thereby placing limitations to the ability to invoke behavior independent of the class hierarchy.

Also, although Simula had nondeterminism, it did not have concurrency.[xxvii]

*Smalltalk-72*

Planner, Simula 67, Smalltalk-72 [Kay 1975; Ingalls 1983] and computer networks had previously used message passing. However, they were too complicated to use as the foundation for a mathematical theory of concurrency. Also they did not address fundamental issues of concurrency.

Alan Kay was influenced by message passing in the pattern-directed invocation of Planner in developing Smalltalk-71 [Kay 1973]. Hewitt was intrigued by Smalltalk-71 but was put off by the complexity of communication that included invocations with many fields including global, sender, receiver, reply-style, status, reply, operator selector, *etc.*

In November 1972 Kay visited MIT and discussed some of his ideas for Smalltalk-72 building on the Logo work of Seymour Papert and the "*little person*" metaphor for computation used for teaching children to program. However, the message passing of Smalltalk-72 was quite complex [Kay 1975]. Code in the language was viewed by the interpreter as simply a stream of tokens.[xxviii] As Dan Ingalls [1983] later described it:[xxix]

> *The first (token) encountered (in a program) was looked up in the dynamic context, to determine the receiver of the subsequent message. The name lookup began with the class dictionary of the current activation. Failing there, it moved to the sender of that activation and so on up the sender chain. When a binding was finally found for the token, its value became the receiver of a new message, and the interpreter activated the code for that object's class.[xxx]*

Thus the message passing model in Smalltalk-72 was closely tied to a particular machine model and programming language syntax that did not lend itself to concurrency. Also, although the system was bootstrapped on itself, the language constructs were not formally defined as objects that respond to **Eval** messages (see discussion below).

The notion of computation has been evolving for a long time. One of the earliest examples was Euclid's GCD algorithm. Next came mechanical calculators of various kinds. These notions were formalized in the Turing Machines, the lambda calculus, etc. paradigm that focused on the "state" of a computation that could be logically inferred from the "previous" state.

*Lambda calculus*

Scott and Strachey [1971] proposed to develop a mathematical semantics for programming languages based on the lambda calculus [Church 1941]. However, the nondeterministic lambda calculus has bounded nondeterminism [Plotkin 1976] and is incapable of implementing concurrency.[xxxi]

*Concurrency*

The invention of digital computers caused a decisive paradigm shift when the notion of an interrupt was invented so that input that arrived asynchronously from outside could be incorporated in an ongoing computation. At first concurrency was conceived using low level machine implementation concepts like threads, locks, channels, cores, queues, *etc.*

The Actor model [Hewitt, Bishop, and Steiger 1973] based computation on message passing.[xxxii] The Actor model has laws that govern privacy and security [Baker and Hewitt 1977].[xxxiii] The break was decisive because asynchronous communication cannot be implemented by Turing machines etc. because the order of arrival of messages cannot be logically inferred. Message passing is the foundation of many-core and client-cloud computing.

*Logic Programming Redux*

Robert Kowalski developed the thesis that "*computation could be subsumed by deduction*" [Kowalski 1988a] that he states was first proposed by Hayes [1973] in the form "*Computation = controlled deduction.*" [Kowalski 1979] This thesis was also implicit in one interpretation of Cordell Green's earlier work [Green 1969].

Kowalski forcefully stated:
> *There is only one language suitable for representing information -- whether declarative or procedural -- and that is first-order predicate logic. There is only one intelligent way to process information -- and that is by applying deductive inference methods.* [Kowalski 1980]

*Actors go beyond Logic Programming*

The gauntlet was officially thrown in *The Challenge of Open Systems* [Hewitt 1985] to which [Kowalski 1988b] replied in *Logic-Based Open Systems* (also see [Davison 2000]). This was followed up with *Guarded Horn clause languages: are they deductive and logical?* [Hewitt and Agha 1988] in the context of the Japanese Fifth Generation Project (see section below). All of this was against Kowalski who stated "*Looking back on our early discoveries, I value most the discovery that computation could be subsumed by deduction.*" [Kowalski 1988a] Kowalski also stated that "*computation could be subsumed by deduction*" [Kowalski 1988a]

According to Hewitt, *et. al.* and contrary to Kowalski and Hayes, computation in general cannot be subsumed by deduction and contrary to the quotation (above) attributed to Hayes computation in general is not controlled deduction. Hewitt and Agha [1991] and other published work argued that mathematical models of concurrency did not determine particular concurrent computations because they make use of arbitration for determining which message is next in the arrival order when multiple messages concurrently. For example Arbiters can be used in the implementation of the arrival order. Since arrival orders are in general indeterminate, they cannot be deduced from prior information by mathematical logic alone. Therefore mathematical logic cannot implement concurrent computation in open systems.

In concrete terms, typically we cannot observe the details by which the arrival order of messages determined. Attempting to do so affects the results and can even push the indeterminacy elsewhere. Instead of observing the internals of arbitration processes, we await outcomes. The reason that we await outcomes is that we have no alternative because of indeterminacy.[xxxiv]

*Computational Representation Theorem*

What does the mathematical theory of Actors have to say about how the representation of the behavior?[xxxv] A closed system is defined to be one that does not receive communications from outside. Actor model theory provides the means to characterize all the possible computations of a closed system in terms of the Computational Representation Theorem [Hewitt 2006]: The denotation $\text{Denote}_S$ of a system $S$ represents all the possible behaviors of $S$ as

$$\text{Denote}_S = \bigsqcup_{i \in \omega} \text{Progression}_S^i(\bot_S)$$

where $\text{Progression}_S$ is an approximation function that takes a set of approximate behaviors to their next stage and $\bot_S$ is the initial behavior of $S$.

In this way, the behavior of S can be mathematically characterized in terms of all its possible behaviors (including those involving unbounded nondeterminism). Although $\text{Denote}_S$ is not an implementation of $S$, it can be used to prove a generalization of the Church-Turing-Rosser-Kleene thesis [Kleene 1943]:

**Enumeration Theorem**: If the primitive Actors of a closed Actor System are effective, then the possible outputs are recursively enumerable.

The upshot is that ***Actor systems can be represented and characterized by logical deduction but cannot be implemented***. Thus, the following practical problem arose:

How can practical programming languages be rigorously defined since the proposal by Scott and Strachey [1971] to define them in terms lambda calculus failed because the lambda calculus cannot implement concurrency?

## *ActorScript™* Programming Language

*A program should not only work, it should also appear to work.*

One solution was to develop a concurrent variant Lisp meta-circular definition [McCarthy, Abrahams, Edwards, Hart, and Levin 1962] that was inspired by Turing's Universal Machine [Turing 1936]. If `exp` is a Lisp expression and `env` is an environment that assigns values to identifiers, then the *procedure* Eval with arguments `exp` and `env` evaluates `exp` using `env`. In the concurrent variant, `Eval(env)` is a *message* that can be sent to `exp` to cause `exp` to be evaluated. Using such messages, modular meta-circular definitions can be concisely expressed in the Actor model for universal concurrent programming languages (e.g. ActorScript [Hewitt 2008f] that is described below).

ActorScript is a general purpose programming language for implementing massive local and nonlocal concurrency. It is differentiated from other concurrent languages by the following:

- Identifiers (names) in the language are referentially transparent, *i.e.*, in a given scope an identifier always refers to the same thing.
- Everything in the language is accomplished using message passing including the very definition of ActorScript itself.
- Functional and Logic Programming are integrated into general concurrent programming.
- Advanced concurrency features such as futures, serializers, sponsors, *etc.* can be defined and implemented without having to resort to low level implementation mechanisms such as threads, tasks, locks, and cores.
- • Binary XML and JSON are data types
- For ease of reading, programming can be displayed using a 2-dimensional textual typography (as is often done in mathematics).

ActorScript attempts to achieve the highest level of performance, scalability, and expressiblity with a minimum of primitives.

> **What went wrong:**
> 1. Nondeterministic global state machines [Dijkstra 1976] failed as a model of concurrent computation. Communicating Sequential Processes[xxxvi] [Hoare 1978] adopted the same model with the result that service could not be formally guaranteed by servers.[xxxvii]
> 2. The thesis that computation is subsumed by deduction failed because concurrent computation could not be implemented.
> 3. The proposal to define the semantics of programming languages in terms of the lambda calculus (a branch of deductive logic) failed because concurrency cannot be implemented in the lambda calculus.
> 4. Concurrent computation was initially conceived in terms of low level machine implementation concepts of threads, locks, channels, queues, *etc.*.
>
> **What was done about it:**
> 1. The Actor model of concurrent computation was developed based on message passing instead of nondeterministic global states.
> 2. A mathematical foundation for concurrent computation was developed based on domain theory [Scott and Strachey 1971, Clinger 1981, Hewitt 2007].
> 3. Universal concurrent programming languages can be modularly defined in terms of themselves using the Actor model.
> 4. The Actor model was developed founding concurrent computation on message passing.

## Japanese Fifth Generation Project (ICOT)

Beginning in the 1970's, Japan became dominant in the DRAM market (and consequently most of the integrated circuit industry). This was accomplished with the help of the Japanese VLSI project that was funded and coordinated mostly by the Japanese government Ministry of International Trade and Industry (MITI) [Sigurdson 1986].

### *Project Inception*

MITI hoped to enlarge this victory by taking over the computer industry with a new Fifth Generation Computing System (FGCS) project (officially named ICOT). However, Japan had come under criticism for "copying" the US. One of the MITI goals for ICOT was to show that Japan could innovate new computer technology and not just copy the United States.

### *Trying to go all the way with the Logic Programming paradigm*

ICOT, strongly influenced by Logic Programming enthusiasts, tried to go all the way with Logic Programming. Kowalski later recalled "*Having advocated LP* [Logic Programming] *as a unifying foundation for computing, I was delighted with the LP focus of the FGCS* [Fifth Generation Computer Systems] *project.*" [Fuchi, Kowalski, Ueda, Kahn, Chikayama, and Tick 1993] By making Logic Programming (which was mainly being developed outside the US) the foundation, MITI hoped that the Japanese computer industry could leapfrog the US. "*The* [ICOT] *project aimed to leapfrog over IBM, and to a new era of advanced knowledge processing applications*" [Sergot 2004]

### *Downfall*

The technical managers at ICOT were aware of some of the pitfalls that had tripped up previous Artificial Intelligence (AI) researchers. So they deliberately avoided calling ICOT an AI Project. Instead they had the vision of an integrated hardware/software system [Uchida and Fuchi 1992]. However, the Logic Programming paradigm turned not to be a suitable foundation because: [Hewitt and Agha 1988].

- *poor modularity:* Logic Programming using clauses was extremely verbose and required the invention of many subsidiary predicates to accomplish simple tasks. In addition there were no organizational principles to use in providing modularity to larger systems.
- *lack of efficiency:* The efficiency of Logic Programming was much less than direct message passing.[xxxviii]

Another problem was that multi-processors found it difficult to compete because at the time single processors were rapidly increasing in speed and connections between multiple processors suffered long latencies.

Thus the overall MITI strategy backfired because and so the Japanese companies refused to productize the ICOT hardware.

However, the architects of ICOT did get some things right:
- The project largely avoided the Mental Agent paradigm
- The project correctly placed tremendous emphasis on research in concurrency and parallelism as an emerging computing paradigm.

| **What went wrong:** |
| --- |
| The way that it used Logic Programming was a principal contributing cause to the failure of ICOT because Logic Programming turned out not to be competitive with message-passing. |
| **What was done about it:** |
| • Japanese companies refused to productize the ICOT architecture.<br>• ICOT languished and then suffered a lingering death. |

## Logic Programming

Arguably, the original paradigm for computation was Logic Programming broadly conceived as "logically inferring computational steps from existing information."

### *Church's Foundation of Logic*

Arguably, Church's *Foundation of Logic* was the first Logic Programming language [Church 1932, 1933].[xxxix] It attempted to avoid the known logical paradoxes by using partial functions and restricting the law of the excluded middle.

The system was very powerful and flexible. Unfortunately, it was so powerful that it was inconsistent [Kleene and Rosser 1935] and consequently the propositional logic was removed, leaving only the functional lambda calculus [Church 1941].

| |
|---|
| **What went wrong:**[xl] <br> A logical system that was developed by Church to be a new foundation for logic turned out to have inconsistencies that could not be removed. |
| **What was done about it:** <br> • Logic was removed from the system leaving the functional lambda calculus, which has been very successful. <br> • Much later a successor system Direct Logic™ [Hewitt 2008f] was developed that overcame these problems of Church's *Foundation of Logic*. (See below.) |

## *McCarthy's Advice Taker*

McCarthy [1958] proposed the Logicist Programme for Artificial Intelligence that included the Advice Taker with the following main features:

2. *There is a method of representing expressions in the computer. These expressions are defined recursively as follows: A class of entities called terms is defined and a term is an expression. A sequence of expressions is an expression. These expressions are represented in the machine by list structures [Newell and Simon 1957].*
3. *Certain of these expressions may be regarded as declarative sentences in a certain logical system which will be analogous to a universal Post canonical system. The particular system chosen will depend on programming considerations but will probably have a single rule of inference which will combine substitution for variables with modus ponens. The purpose of the combination is to avoid choking the machine with special cases of general propositions already deduced.*
4. *There is an immediate deduction routine which when given a set of premises will deduce a set of immediate conclusions. Initially, the immediate deduction routine will simply write down all one-step consequences of the premises. Later, this may be elaborated so that the routine will produce some other conclusions which may be of interest. However, this routine will not use semantic heuristics; i.e., heuristics which depend on the subject matter under discussion.*
5. *The intelligence, if any, of the advice taker will not be embodied in the immediate deduction routine. This intelligence will be embodied in the procedures which choose the lists of premises to which the immediate deduction routine is to be applied.*
6. *The program is intended to operate cyclically as follows. The immediate deduction routine is applied to a list of premises and a list of individuals. Some of the conclusions have the form of imperative sentences. These are obeyed. Included in the set of imperatives which may be obeyed is the routine which deduces and obeys.*

McCarthy summarized that in the Advice Taker, "*the procedures will be described as much as possible in the language itself and, in particular, the heuristics are all so described*" which is what Logic Programming is all about.

| **What went wrong:** |
|---|
| • The imperative sentences deduced by the Advice Taker could have impasses in the following forms: |
|     • *lapses* in which no imperative sentences were deduced |
|     • *conflicts* in which inconsistent sentences were deduced. |
| • The immediate deduction routine of the Advice Taker was extremely inefficient |
| **What was done about it:** |
| • McCarthy, *et al.,* developed Lisp (one of the world's most influential programming languages) in order to implement ideas in the Advice Taker and other AI systems. Using Lisp, Minsky, *et al.* developed a procedural approach to AI [Minsky 1968] building on the work of [Newell and Simon 1956, Gelernter 1959, *etc.*]. |
| • McCarthy changed the focus of his research to solving epistemological problems of Artificial Intelligence |
| • The Soar architecture was developed to deal with impasses [Laird, Newell, and Rosenbloom 1987]. |

## *Uniform Proof Procedures based on Resolution*

John Alan Robinson [1965] developed a deduction method called resolution that was proposed as a uniform proof procedure. Resolution required converting everything to clausal form and then used a method analogous to modus ponens to attempt to obtain a proof by contradiction by adding the clausal form of the negation of the theorem to be proved.

The first use of Resolution was in computer programs to prove mathematical theorems and in the synthesis of simple sequential programs from logical specifications [Wos 1965; Green 1969; Waldinger and Lee 1969; Anderson and 1970; 1971, *etc.*]. In the resolution uniform proof procedure theorem proving paradigm, the use of procedural knowledge was considered to be "*cheating*" [Green 1969].

| **What went wrong:** |
|---|
| • Using resolution as the only rule of inference is problematical because it hides the underlying structure of proofs by comparison with Natural Deduction (*e.g.* [Fitch 1952]). |
| • It proved to be impossible to develop efficient enough uniform proof procedures for practical domains.[xli] |
| • Using proof by contradiction is problematical because the axiomatizations of all practical domains of knowledge are inconsistent in practice. And proof by contradiction is not a sound rule of inference for inconsistent systems. |
| **What was done about it:** |
| • The *Procedural Embedding of Knowledge* paradigm [Hewitt 1971] based on the invocation of plans from goals and assertions was developed as an alternative to *Resolution Uniform Proof Procedure* paradigm. (See below.) |
| • Direct inference (such as in Direct Logic [Hewitt 2008f]) was developed to isolate inconsistencies during reasoning. (See section below.) |

## *Planner*

The two major paradigms for constructing information integration systems were procedural and logical. The procedural paradigm was epitomized by using Lisp [McCarthy *et al.* 1962; Minsky, *et al.* 1968] recursive procedures operating on list structures. The logical paradigm was epitomized by uniform resolution theorem provers [Robinson 1965].

Planner [Hewitt 1969] was a kind of hybrid between the procedural and logical paradigms. An implication of the form (P *implies* Q) was procedurally interpreted as follows:[xlii]

- *when assert* P**,** *assert* Q
- *when goal* Q**,** *goal* P
- *when assert* (*not* Q)**,** *assert* (*not* P)
- *when goal* (*not* P)**,** *goal* (*not* Q)

Planner was the first programming language based on the pattern-directed invocation of procedural plans from assertions and goals. The development of Planner was inspired by the work of Karl Popper [1935, 1963], Frederic Fitch [1952], George Polya [1954], Allen Newell and Herbert Simon [1956], John McCarthy [1958, *et. al.* 1962], and Marvin Minsky [1968]. ***Planner represented a rejection of the resolution uniform proof procedure paradigm.***

Computers were expensive. They had only a single slow processor and their memories were very small by comparison with today. So Planner adopted some efficiency expedients including the following:

- Backtracking [Golomb and Baumert 1965] was adopted to economize on the use of time and storage by working on and storing only one possibility at a time in exploring alternatives.
- A unique name assumption was adopted to save space and time by assuming that different names referred to different objects. For example names like Peking and Beijing were assumed to refer to different objects.
- A closed world assumption could be implemented by conditionally testing whether an attempt to prove a goal exhaustively failed. Later this capability was given the misleading name "negation as failure" because for a goal *G* it was possible to say: "if attempting to achieve *G* exhaustively fails then assert Not *G*.

A subset called Micro-Planner was implemented by Gerry Sussman, Eugene Charniak and Terry Winograd. Micro-Planner was used in Winograd's natural-language understanding program SHRDLU [Winograd 1971], Eugene Charniak's story understanding work, work on legal reasoning [McCarty 1977], *etc*. This generated a great deal of excitement in the field of AI. Since Micro-Planner was embedded in Lisp, applications used two different syntaxes and so lacked a certain degree of elegance. In fact, after Hewitt's lecture at IJCAI'71, Allen Newell rose from the audience to remark on the "Baroque" syntax! However, variants of this syntax persist to this day.

| |
|---|
| **What went wrong:** |
| 1. Although pragmatically useful at the time Planner was developed, the efficiency expedients (backtracking, unique name assumption, and closed world assumption) proved to be rigid and inexpressive. |
| 2. Planner had a single global data base that was not modular or scalable. |
| 3. Although pragmatically useful for interfacing with the underlying Lisp system, the syntax used in micro-Planner applications was not a pretty sight. |
| **What was done about it:** |
| 1. Concurrency based on message passing was developed as an alternative to backtracking. [Hewitt, Bishop, and Steiger 1973] |
| 2. *QA4* [Rulifson, Derksen, and Waldinger 1973] developed a hierarchical context system to modularize the data base. Contexts were later generalized in Direct Logic [Hewitt 2008f] (see below). |
| 3. Prolog [Kowalski 1974, Colmerauer and Roussel 1996] was basically a subset of Planner that restricted programs to clausal form using backward chaining (*i.e.* no forward chaining) and no ability to define functions. Consequently Prolog had a simpler syntax than Planner. However, the simpler syntax came at the cost of expressive power including not being able to make assertions and the inability to directly say that an assertion is false. (See [Hewitt 2008c] for further information.) |

## Acknowledgements

Alonzo Church, Alain Colmerauer, Ted Elcock, Scott Fahlman, Solomon Feferman, Frederic Fitch, Cordell Green, Pat Hayes, Stephen Kleene, Bill Kornfeld, Robert Kowalski, John McCarthy, Drew McDermott, Marvin Minsky, Alan Robinson, Philippe Roussel, John Barkley Rosser, Jeff Rulifson, Erik Sandewall, Dana Scott, Christopher Strachey, Gerry Sussman, Alan Turing, Richard Waldinger, etc. deserve a lot of credit for contributing to the development of Logic Programming. At the same time, the term "logic programming" (like "functional programming") is highly descriptive and should mean something. Over the course of history, the term "functional programming" has grown more precise and technical as the field has matured. Logic Programming should be on a similar trajectory. Accordingly, "Logic Programming" should have a more precise characterization, *e.g.,* "*the logical inference of computational steps*".

Today we know much more about the strengths and limitations of Logic Programming than in the late 1960's. For example, Logic Programming is not computationally universal and is strictly less general than the Procedural Embedding of Knowledge paradigm [Minsky *et al*. 1968; Hewitt 1971]. Logic Programming and Functional Programming will both be very important for concurrent computation. Although neither one by itself (or even both together) can do the whole job, what can be done is extremely well suited to massive concurrency.

At AAAI'08, conversations with Bruce Buchanan, Mehmet Göker, Ben Kuipers, Erik Sandewall, Dan Shapiro, Reid Smith, and others were very helpful. The Stanford Logic group led by Michael Genesereth has provided a supportive environment for the further development of some of the ideas. Elihu M. Gerson, Erik Sandewall, and Dan Shapiro made extensive suggestions for improving this paper. Richard Waldinger provided a suggestion on how to better characterize Logic Programming. Conversations with Pat Helland were very helpful in developing the blob storage service described in this paper. Jeremy Forth made helpful comments and suggested including the section on the future of Logic Programming. Peter Neumann proposed that I develop a better ending for the paper. An anonymous referee of CACM corrected some typos. Peter de Jong and Fanya S. Montalvo provided extensive comments and suggestions. Participant

# End Notes

[i] It is not possible to guarantee the consistency of information because consistency testing is recursively undecidable even in logics much weaker than first order logic. Because of this difficulty, it is impractical to test whether information is consistent.

[ii] Consequently iDescriber makes use of direct inference in Direct Logic to reason more safely about inconsistent information because it omits the rules of classical logic that enable every proposition to be inferred from a single inconsistency.

[iii] Each iPhrase is an iDescriber.

[iv] See [Karp, Stiegler, and Close 2009] and [Stiegler 2009] for important work that can be applied.

[v] Some of the material in this section is derived from [Russell and Hewitt 2010].

[vi] Some of the material in this section and the following one on Mental Agents was published in [Hewitt 2009a].

[vii] Organizations of Restricted Generality™

[viii] In contrast with the Mental Agent paradigm, iOrgs can have people that are tightly integrated with information technology that enables them to function organizationally. Humans are integral to the operation of iOrgs since they continually (re)design, debug, monitor, (re)install, (re)boot, and so forth. In many cases, humans take part in the decisions. For example, in credit card verification iOrg, a human might review a particular transaction in the course of an iOrg processing it (see [Licklider 1960]).

[ix] This is a form of two-factor access control: *Warrants* and *iOrgs*. Warrants express the authority to take specified actions and iOrgs specify the organizational authority ranging from an individual role to a whole organization.

[x] Similar sentiments can be found in independent work by [Finkelstein, Brendle, and Jacobs 2009], [Helland and Campbell 2009], and [Armbrust, et. al. 2009].

[xi] Paraconsistency (name coined by Francisco Miró Quesada in 1976 [Priest 2002, pg. 288]) was developed to deal with inconsistent theories. The idea of paraconsistent logic is to be able to make inferences from inconsistent information without being able to derive all propositions. Paraconsistency is a much weaker property than "*direct inference*" in Direct Logic.

The most extreme form of paraconsistent mathematics is *dialetheism* [Priest and Routley 1989] which maintains that there are true inconsistencies in mathematics itself *e.g.,* the Liar Paradox. However, mathematicians (starting with Euclid) have worked very hard to make their theories consistent and inconsistencies have not been an issue for most working mathematicians. As a result:
- Since inconsistency was not an issue, mathematical logic focused on the issue of truth and a model theory of truth was developed by Dedekind (1888), Löwenheim (1915), Skolem (1920), Gödel (1930), Tarski and Vaught (1957), and Hodges (2006). More recently there has been work on the development of an unstratified logic of truth [Leitgeb 2007, Feferman 2007a].
- Paraconsistent logic somewhat languished for lack of subject matter. The lack of subject matter resulted in paraconsistent systems that were for the most part so awkward as to be unused for mathematical practice.

Consequently mainstream logicians and mathematicians have tended to shy away from paraconsistency.

***One of the achievements of Direct Logic is the development of an inconsistency robust direct inference system with mathematical induction that does minimal damage to traditional natural deductive logical reasoning.***

Paraconsistent logics have not been satisfactory for the purposes of Software Engineering because of their many seemingly arbitrary variants and their idiosyncratic inference rules and notation. For example (according to Priest [2006]), most paraconsistent and relevance logics rule out Disjunctive Syllogism (($\Phi \vee \Psi$), $\neg \Phi$ ⊢ $\Psi$). However, Disjunctive Syllogism seems entirely natural for use in Software Engineering!

[xii] Direct inference is defined differently in Direct Logic from probability theory [Kyburg and Teng 2001], which refers to "direct inference" of frequency in a reference class (the most specific class with suitable frequency knowledge) from which other probabilities are derived.

[xiii] In theory *Boston*, a weekday at 5PM infers a traffic jam.

[xiv] In theory *Boston*, no traffic jam.

[xv] If the contrapositive is intended in Direct Logic, then A1 would be ⊢$_{Boston}$ **WeekdayAt5PM** ⇨ **TrafficJam** where ⇨ is implication.

[xvi] Many in the nonmonotonic community have omitted contraposition from rules, *e.g.,* [Reiter 1980; Prakken and Sartor 1996; Caminda 2008]. According to [Ginsberg 1994 pg. 16]: "*although almost all of the symbolic approaches to nonmonotonic reasoning do allow for the strengthening of the antecedents of default rules, many of them do not sanction contraposition of these rules.*"

In Direct Logic, nonmonoticity can be accommodated by using another theory *BostonWithoutA1* that is derived from *Boston* by omitting A1 so that there is no inconsistency from the following:

⊢$_{BostonWithoutA1}$ ¬**TrafficJam**

⊢$_{BostonWithoutA1}$ **WeekdayAt5PM**

Instead, the theory *BostonWithoutA1* might have the following:

**WeekdayAt5PM**, ¬**Holiday** ⊢$_{BostonWithoutA1}$ **TrafficJam**

[xvii] The same issue affects probabilistic (fuzzy logic) systems. Suppose (as above) the probability of TrafficJam is 0 and the probability of (TrafficJam given WeekdayAt5PM) is 1. Then the probability of WeekdayAt5PM is 0. Varying the probability of TrafficJam doesn't change the principle involved because the probability of WeekdayAt5PM will always be less than or equal to the probability of TrafficJam.

[xviii] Boolean propositions use only the connectives for conjunction, disjunction, implication, and negation.

[xix] In this way Direct Logic differs from Relevance Logic because Boolean Relevance Logic is recursively undecidable

[xx] The probability is **1** for **TrafficJam** given **WeekdayAt5PM**.

[xxi] Varying **P**(**TrafficJam**) doesn't change the principle involved because ⊢$_{Boston}$ **P**(**WeekdayAt5PM**) ≦ **P**(**TrafficJam**)

[xxii] Ontology Web Language that is part of the Semantic Web. According to Parastatidis, Viegas, and Hey [2009]:
*We make a distinction between the general approach of computing based on semantic technologies (machine learning, neural networks, ontologies, inference and so forth) and the "Semantic Web" as described in [Berners-Lee et.al. 2001] and [Shadbolt et. al. 2006] and which is the term used to refer to a specific ecosystem*

*of technologies, like RDF and OWL. The Semantic Web has gained a lot of attention lately, bringing more awareness of the importance of semantics. However, we consider the Semantic Web technologies to be just some of the many tools at a disposal when building semantics-based solutions.*

[xxiii] Some theories are also called ontologies.

[xxiv] There has been some work on developing inconsistency robust reasoners for OWL-like languages outside the W3C specifications including PION [Zhisheng Huang, van Harmelen, and ten Teije 2005] and ParOWL [Ma, Hitzler and Lin 2006]. Unfortunately, both lack some means of reasoning that are important in iDescriber applications, *e.g.*, roundtripping [Hewitt 2008e].

[xxv] Fortunately, the limitations of the W3C specifications can be overcome in a way that substantially preserves work using them so that it doesn't have to be completely redone. Horrocks [2008] has a recent overview of OWL.

[xxvi] Consequently in Simula-76 there was no required locality of operations unlike the laws for locality in the Actor mode [Baker and Hewitt 1977].

[xxvii] The ideas in Simula became widely known by the publication of [Dahl and Hoare 1972] at the same time that the Actor model was being invented to formalize concurrent computation using message passing [Hewitt, Bishop, and Steiger 1973].

[xxviii] Subsequent versions of the Smalltalk language largely followed the path of using the virtual methods of Simula 67 in the message passing structure of programs. However Smalltalk-72 made primitives such as integers, floating point numbers, etc. into objects. The authors of Simula 67 had considered making such primitives into objects but refrained largely for efficiency reasons. Java at first used the expedient of having both primitive and object versions of integers, floating point numbers, etc. The C# programming language (and later versions of Java, starting with Java 1.5) adopted the more elegant solution of using boxing and unboxing, a variant of which had been used earlier in some Lisp implementations.

[xxix] The Smalltalk system went on to become very influential, innovating in bitmap displays, personal computing, the class browser interface, and many other ways. Meanwhile the Actor efforts at MIT remained focused on developing the science and engineering of higher level concurrency.

See Briot [1988] for ideas that were developed later on how to incorporate some kinds of Actor concurrency into later versions of Smalltalk.

[xxx] According to the Smalltalk-72 Instruction Manual [Goldberg and Kay 1976]:
> There is not one global message to which all message "fetches" (use of the Smalltalk symbols eyeball, ◁; colon, ⁞, and open colon, ⸰) refer; rather, messages form a hierarchy which we explain in the following way-- suppose I just received a message; I read part of it and decide I should send my friend a message; I wait until my friend reads his message (the one I sent him, not the one I received); when he finishes reading his message, I return to reading my message. I can choose to let my friend read the rest of my message, but then I can not get the message back to read it myself (note, however, that this can be done using the Smalltalk object *apply* which will be discussed later). I can also choose to include permission in my message to my friend to ask me to fetch some information from my message and to give that in information to him (accomplished by including: or ⸰ in the message to the friend). However, anything my friend fetches, I can no longer have. In other words,

1) An object (let's call it the CALLER) can send a message to another object (the RECEIVER) by simply mentioning the RECEIVER's name followed by the message.
2) The action of message sending forms a stack of messages; the last message sent is put on the top.
3) Each attempt to receive information typically means looking at the message on the top of the stack.

4) The RECEIVER uses the eyeball, ◀ the colon, :, and the open colon, ○, to receive information from the message at the top of the stack.
5) When the RECEIVER completes his actions, the message at the top of the stack is removed and the ability to send and receive messages returns to the CALLER. The RECEIVER may return a value to be used by the CALLER.
6) This sequence of sending and receiving messages, viewed here as a process of stacking messages, means that each message on the stack has a CALLER (message sender) and RECEIVER (message receiver). Each time the RECEIVER is finished, his message is removed from the stack and the CALLER becomes the current RECEIVER. The now current RECEIVER can continue reading any information remaining in his message.
7) Initially, the RECEIVER is the first object in the message typed by the programmer, who is the CALLER.
8) If the RECEIVER's message contains an eyeball, ◀; colon, :, or open colon, ○, he can obtain further information from the CALLER's message. Any information successfully obtained by the RECEIVER is no longer available to the CALLER.
9) By calling on the object *apply,* the CALLER, can give the RECEIVER the right to see all of the CALLER's remaining message. The CALLER can no longer get information that is read by the RECEIVER; he can, however, read anything that remains after the RECEIVER completes its actions.
10) There are two further special Smalltalk symbols useful in sending and receiving messages. One is the keyhole, ⌑, that lets the RECEIVER "peek" at the message. It is the same as the ○ except it does not remove the information from the message. The second symbol is the hash mark, #, placed in the message in order to send a reference to the next token rather than the token itself.

xxxi In 1975, Sussman and Steele took an interest in Actors. They noticed some similarities between Actor customers and continuations introduced by [Reynolds 1972] using a primitive called *escape*. They called their variant of *escape* by the name "*call with current continuation.*" Unfortunately, general use of *escape* is not compatible with usual hardware stack disciple introducing considerable operational inefficiency. Also, using *escape* can leave customers stranded [Hewitt 2009c]. Consequently, use of *escape* is generally avoided these days and exceptions are used instead so that clean up can be performed.

Sussman and Steele 1975] mistakenly concluded "*we discovered that the 'Actors' and the lambda expressions were identical in implementation.*" The actual situation is that the lambda calculus is capable of expressing some kinds of sequential and parallel control structures but, in general, *not* the concurrency expressed in the Actor model.

The lambda calculus includes the following limitations:
- Message arrival ordering cannot be implemented
- Actors cannot be implemented whose behavior evolves with time
- The lambda calculus does not have exceptions and consequently neither does Scheme
- Modeling customers as continuation functions led to the hanging customer issue [Hewitt 2009c]

On the other hand, the Actor model is capable of expressing everything in the lambda calculus and more.

xxxii Process calculi (*e.g.* [Milner 1993; Cardelli and Gordon 1998]) are closely related to the Actor model. There are many similarities between the two approaches, but also several differences (some philosophical, some technical):
- There is only one Actor model (although it has numerous formal systems for design, analysis, verification, modeling, etc.); there are numerous process calculi, developed for reasoning about a variety of different kinds of concurrent systems at various levels of detail (including calculi that incorporate time, stochastic transitions, or constructs specific to application areas such as security analysis).
- The Actor model was inspired by the laws of physics and depends on them for its fundamental axioms, i.e. physical laws (see Actor model theory); the process calculi were originally inspired by algebra [Milner 1993].
- Semantics of the Actor model is based on message orderings in the Computational Representation Theorem. Semantics of process calculi are based on structural congruence in various kinds of bisimulations and equivalences.
- Computational objects in process calculi are anonymous, and communicate by sending messages either through named channels (synchronous or asynchronous), or via ambients (which can also be used to model channel-like communications [Cardelli and Gordon 1998]). In contrast, Actors in the Actor model possess an identity, and communicate by sending messages to the mailing addresses of other Actors (this style of communication can also be used to model channel-like communications).

For example, communication in the π–calculus [Milner 1993] takes the following form:
- *input:* u(x).P is a process that gets a message from a communication channel named u before proceeding as P, binding the message received to the identifier x. In ActorScript, this can be modeled as follows:
  *let* (x=u.*get*;) P
- *output:* ū<m>.P is a process that puts a message m on communication channel u before proceeding as P. In ActorScript, this can be modeled as follows: {u.*put* (m); P}

The rest of the π-calculus can be modeled using a two-phase commit protocol [Knabe 1992; Reppy, Russo, and Xiao 2009].

[xxxiii] These laws can be enforced by a proposed extension of the X86 architecture that will support the following operating environments:
- CLR and extensions (Microsoft)
- JVM (Sun, IBM, Oracle, SAP)
- Dalvik (Google)

Many-core 3-D through-silicon via architecture has made the above extension necessary in order to provide the following:
- concurrent nonstop automatic storage reclamation (garbage collection) and relocation to improve efficiency,
- prevention of memory corruption that otherwise results from programming languages like C and C++ using thousands of threads in a process,
- nonstop migration of iOrgs (while they are in operation) within a computer and between distributed computers.

[xxxiv] Although [Goldin and Wegner 2008] may seem superficially similar, it unfortunately failed to comprehend previous publications on the Actor model (*e.g.* [Hewitt, Bishop and Steiger 1973], [Hewitt 1977], and [Hewitt and Agha 1988]).

[xxxv] Other models of concurrency can be modeled using a two-phase commit protocol [Knabe 1992].

[xxxvi] CSP differed from the Actor model in the following respects:
- *The concurrency primitives of CSP were input, output, guarded commands, and parallel composition* whereas the Actor model is based on asynchronous one-way messaging.
- *The fundamental unit of execution was a sequential process* in contrast to the Actor model in which execution was fundamentally concurrent. Sequential execution is problematical because multi-processor computers are inherently concurrent.
- *The processes had a fixed topology of communication* whereas Actors had a dynamically changing topology of communications. Having a fixed topology is problematical because it precludes the ability to dynamically adjust to changing conditions.
- *The processes were hierarchically structured using parallel composition* whereas Actors allowed the creation of non-hierarchical execution using futures [Baker and Hewitt 1977]. Hierarchical parallel composition is problematical because it precludes the ability to create a process that outlives its creator. Also message passing is the fundamental mechanism for generating parallelism in the Actor model; sending more messages generates the possibility of more parallelism.

- *Communication was synchronous* whereas Actor communication was asynchronous. Synchronous communication is problematical because the interacting processes might be far apart.
- *Communication was between processes* whereas in the Actor model communications are one-way to Actors. Synchronous communication between processes is problematical by requiring a process to wait on multiple processes.
- *Data structures consisted of numbers, strings, and arrays* whereas in the Actor model data structures were Actors. Restricting data structures to numbers, strings, and arrays is problematical because it prohibits programmable data structures.
- *Messages contain only numbers and strings* whereas in the Actor model messages could include the addresses of Actors. Not allowing addresses in messages is problematical because it precludes flexibility in communication because there is no way to supply another process with the ability to communicate with an already known process.

[xxxvii] Hoare [1985] developed a revised version of CSP with unbounded nondeterminism [Roscoe 2005].

[xxxviii] ICOT had to deal with concurrency and consequently developed concurrent programming languages based on clauses that were loosely related to logic [Shapiro 1989]. However, it proved difficult to implement clause invocation in these languages as efficiently as procedure invocation in object-oriented programming languages. Simula-67 originated a hierarchical class structure for objects so that message handling procedures (methods) and object instance variables could be inherited by subclasses. Ole-Johan Dahl [1967] invented a powerful compiler technology using dispatch tables that enabled message handling procedures in subclasses of objects to be efficiently invoked. The combination of efficient inheritance-based procedure invocation together with class libraries and browsers (pioneered in Smalltalk) was better than the slower pattern-directed clause invocation of the FGCS programming languages. Consequently, the ICOT programming languages never took off and instead concurrent object-oriented message-passing languages like Java and C# became the mainstream.

[xxxix] Of course, this was back when computers were humans!

[xl] In research, things invariably go wrong. Typically, no is to blame. Often, participants disagree about what if anything is wrong and what to do about it. The fundamental lesson is humility: "*We don't know much. And some of it is wrong. But we don't know which parts!*"

[xli] In other words, taking a first order axiomatization of a large practical domain, converting it to clausal form, and then using a uniform resolution proof procedure was found to be so wildly inefficient that answers to questions of interest could not be found even though they were logically entailed.

[xlii] This turned out later to have a surprising connection with Direct Logic. See the Two-Way Deduction Theorem below.